\documentstyle[proceedings, epsf]{crckapb} 

\makeatletter                                            
\@ifundefined{epsfbox}{\@input{epsf.sty}}{\relax}        
\def\plotone#1{\centering \leavevmode                    
\epsfxsize=\columnwidth \epsfbox{#1}}                    
\def\plotone_reduction#1#2{\centering \leavevmode        
\epsfxsize=#2\columnwidth \epsfbox{#1}}                  
\makeatother                                             
\begin{opening}
\title{3D Finite Volume Simulation of Accretion Discs with Spiral Shocks}
\author{Makoto Makita}
\author{Takuya Matsuda}
\institute{Department of Earth and Planetary Sciences, Kobe University, 
Kobe 657-8501, Japan}
\end{opening}
\begin{document}
\section{Introduction}
   One of the models of accretion discs in a close binary system is
the spiral shock model. This model was first proposed by one of the
present authors (Sawada, Matsuda \& Hachisu 1986a, b, Sawada et
al. 1987). A number of authors have confirmed since then that spiral
shocks appear in two-dimensional discs. In the case of 3D, Sawada \&
Matsuda (1992) used a Roe scheme to calculate the case of $\gamma=1.2$
with mass ratio of unity. They found the existence of spiral shocks,
but their calculation was done up to only half a revolution
period. Yukawa, Boffin \& Matsuda (1997), who made numerical
simulations using SPH with mass ratio of unity, also demonstrated that
spiral shocks existed in the case of $\gamma=1.2$. 

   The purpose of the present paper is to perform 2D and 3D finite volume 
numerical calculations with higher resolution up to enough time to confirm 
the existence of spiral shocks. 

\section{Calculations and Conclusions}
   We considered a binary system composed of a mass-accreting primary star 
and a mass-losing secondary star. The mass ratio of the mass-losing star to 
the mass-accreting star was one. We set a numerical grid centered at the 
primary star and just touching the inner Lagrangian point L1. In addition, 
we used Cartesian coordinates having $200\times200$ grid points in 2D and 
$200\times200\times50$ grid points in 3D, respectively. Finally, we assumed 
a polytropic gas with constant specific heat ratio $\gamma$.

\begin{figure}[hbt]
\begin{center}
\leavevmode 
\epsfxsize=41mm
\epsfbox{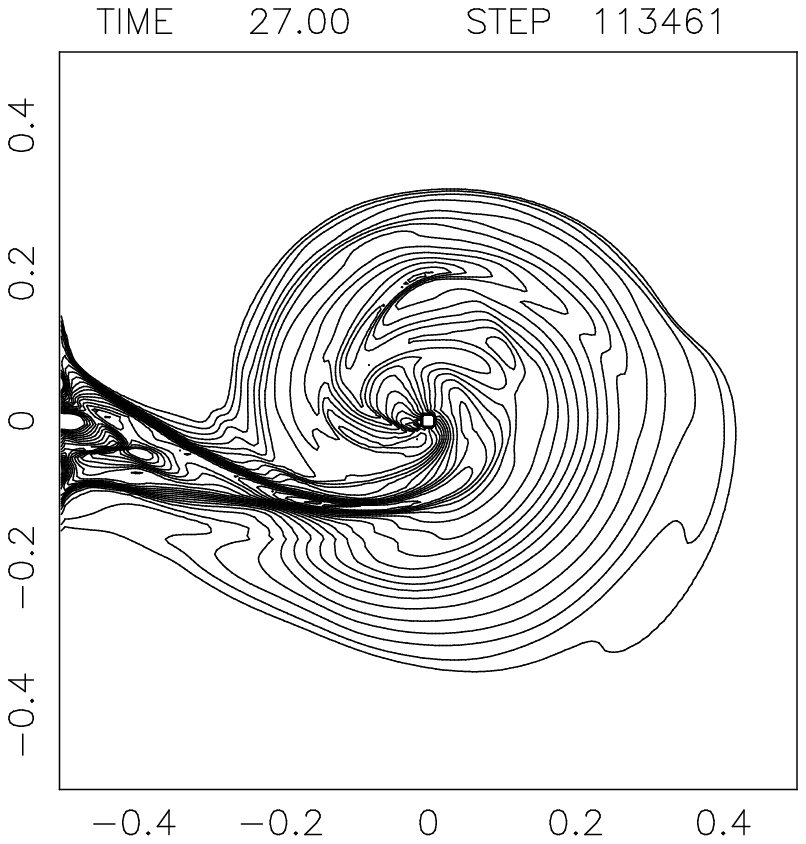}
\hspace{5mm}
\epsfxsize=41mm
\epsfbox{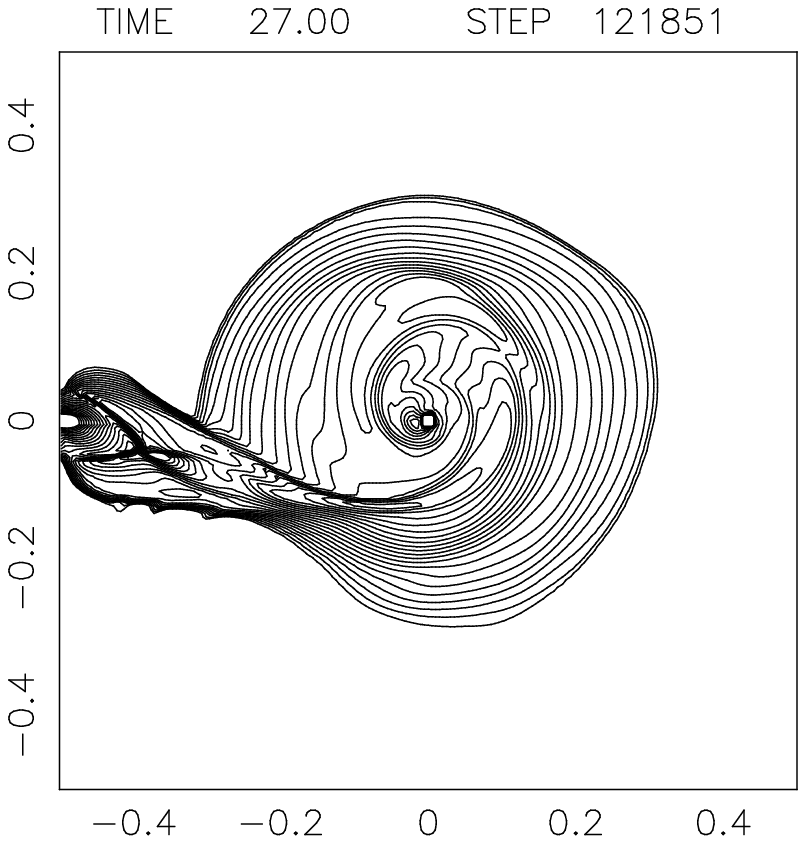}
\end{center}
\caption[dxy]{Density contours in the orbital plane at $t=27$. 
Left: the case of $\gamma=1.2$, Right: the case of $\gamma=1.01$.} 
\label{fig:densex-y}
\end{figure}

   In this paper, we examined the cases of $\gamma=$1.2, 1.1, 1.05, and 
1.01. Symmetry about the orbital plane was assumed in our 3D calculations. 
We used the Simplified Flux vector Splitting (SFS) scheme (Jyounouti et al. 
1993; Shima \& Jyounouti 1994). In 2D calculations, we found that the 
smaller $\gamma$ was, the more tightly the spiral wound. The density 
contours in the orbital plane in the cases of $\gamma=1.2$ and 
$\gamma=1.01$ of our 3D simulations are presented in Fig.\ref{fig:densex-y}. 
These figures show the existence of spiral shocks in the cases of $\gamma=1.2$
 as well as in all the other cases.
However, they do not show as clear a difference in spiral structure as they 
do in our 2D calculations. (Other results can be seen in CD-ROM).

\vspace{-3mm}

Steeghs, Harlaftis \& Horne (1997) found the first convincing evidence of 
spiral structure in accretion discs observationally. We have now succeeded 
in reproducing their results using our numerical calculations. (Please see 
Matsuda et al. in this volume).

\section*{Acknowledgments}
The 2D calculations were mainly performed on VX/1R at the NAOJ.
The 3D calculations were performed on SX-4 at IPC at Kobe University.

\end{document}